\documentclass[12pt]{article}

\usepackage{amssymb}
\newcommand{\h}{{\cal H}}

\newcommand{\Complex}{\mathbb C}

\newcommand{\V}{\ensuremath{{\cal{V}} \, }}

\newcommand{\E}{\ensuremath{{\cal{E}} \, }}
\newcommand{\G}{\ensuremath{{\cal{G}} \, }}

\newcommand{\N}{\ensuremath{{\cal{N}} \, }}
\newcommand{\D}{\ensuremath{{\cal{D}} \, }}

\newcommand{\C}{\ensuremath{{\cal{C}} \, }}

\newcommand{\M}{\ensuremath{{\cal{M}} \, }}

\newcommand{\F}{\ensuremath{{\cal{F}} \, }}

\newcommand{\Real}{\ensuremath{{\mathbb{R}} \, }}

\newcommand{\cP}{\ensuremath{{\cal{P}} \, }}

\newcommand{\U}{\ensuremath{{\cal{U}} \, }}
\newcommand{\cS}{\ensuremath{{\cal{S}} \, }}

\newcommand{\p}{\partial}

\newcommand{\ie}{\emph{i.e.},}
\newcommand{\eg}{\emph{eg.},}

 \newtheorem{thm}{Theorem}[subsection]
 \newtheorem{cor}[thm]{Corollary}
 \newtheorem{lem}[thm]{Lemma}
 \newtheorem{prop}[thm]{Proposition}
 \newtheorem{defn}[thm]{Definition}
 
\begin{document}

\begin{titlepage}

\title{\bf{BRST Quantisation of Histories Electrodynamics}}
\author{Duncan Noltingk\footnote{e-mail d.noltingk@ic.ac.uk}\\
        \\
        Blackett Laboratory\\
        Imperial College\\
        Prince Consort Road\\
        London SW7 2BZ}

\maketitle


\begin{abstract}
This paper is a continuation of earlier work where a classical
history theory of pure electrodynamics was developed in which the
the history fields have \emph{five} components. The extra
component is associated with an extra constraint, thus enlarging
the gauge group of histories electrodynamics. In this paper we
quantise the classical theory developed previously by two
methods. Firstly we quantise the reduced classical history space,
to obtain a reduced quantum history theory. Secondly we quantise
the classical BRST-extended history space, and use the BRST charge
to define a `cohomological' quantum history theory. Finally we
show that the reduced history theory is isomorphic (as a history
theory) to the cohomological history theory.
\end{abstract}

\end{titlepage}

\section{Introduction}

The history projection operator (HPO) approach to consistent
histories was inspired by Isham \cite{Ish94}, and developed by
Isham and collaborators \cite{IshLin95, IshLin98}. An HPO theory
is concerned with projection operators on the quantum history
Hilbert space $\E$ which represent propositions about the entire
\emph{history} of the system under consideration. This should be
contrasted with standard quantum logic which is concerned with
propositions about the system at a particular instant of time.
The quantum history theories with which this paper will be
concerned are described by a pair $(P(\E),\D)$ where $P(\E)$ is
the lattice of projection operators on the Hilbert space $\E$, and
$\D$ is the space of decoherence functionals. A decoherence
functional is a map $d : P(\E) \times P(\E) \rightarrow \Complex$
that satisfies the following conditions:

\smallskip

1. \emph{Hermiticity}: $d(\alpha, \beta) = d(\beta, \alpha)^*$ for
all $\alpha, \beta \in P(\E)$.

2. \emph{Positivity}: $d(\alpha,\alpha) \geq 0$ for all $\alpha
\in P(\E)$.

3. \emph{Null Triviality}: $d(0,\alpha) = 0$ for all $\alpha \in
P(\E)$.

4. \emph{Additivity}: if $\alpha \perp \beta$ then, $d(\alpha
\oplus \beta,\gamma) = d(\alpha,\gamma) + d(\beta,\gamma)$,

for all $\gamma \in P(\E)$.

5. \emph{Normalisation}: $d(1,1)=1$.

\smallskip

The off-diagonal components of the decoherence functional
represent the `quantum interference' between histories, while the
diagonal components are interpreted as the probability that a
particular history `occurs'.

In this paper we construct a quantum history theory of pure
electrodynamics by two methods. In section 2 we quantise the
reduced classical history space to obtain a reduced history theory
$(P(\E^{red}),d^{red})$. In section 3 we augment the history
fields with ghost fields and quantise the extended theory to
obtain a representation of the extended algebra on the
BRST-extended history space $\E$. We then define $H^*(\Omega)$,
the projection operator cohomology, and show that it has a
natural lattice structure. In section 4 we define $\D_{gf}$, the
space of gauge-fixed decoherence functionals, and show that each
gauge-fixed decoherence functional induces a functional
$\tilde{d}$ on $H^*(\Omega)$. Our main result is to show that the
cohomological history theory $(H^*(\Omega),\tilde{d})$ is
isomorphic (as a history theory) to the reduced history theory
$(P(\E^{red}),d^{red})$.

\newpage

\section{Radiation Gauge Quantisation}

\subsection{Preliminaries}

The quantum history space arises as the representation space of a
certain Lie group, the \emph{history group}, and the associated
Lie algebra is the histories analogue of the canonical commutation
relations. In the histories approach to scalar field theory
proposed by Savvidou \cite{Sav01a}, there is an inequivalent
representation of the history group for each Lorentzian foliation
of space-time. By a Lorentzian foliation of space-time we mean a
foliation in which each leaf is a space-like hyperplane. The
\emph{Schr\"odinger} picture fields satisfy the \emph{covariant}
history algebra
\begin{eqnarray}
\left[ \hat{\phi}_n(X) , \hat{\phi}_n(X') \right] &=& 0 \\
\left[ \hat{\pi}_n(X) , \hat{\pi}_n(X') \right] &=& 0 \\
\left[ \hat{\phi}_n(X) , \hat{\pi}_n(X') \right] &=& i\hbar
\delta^{(4)}(X-X'),
\end{eqnarray}
where $n$ is a future pointing time-like unit vector labelling a
particular Lorentzian foliation. The fields are genuine space-time
fields under the action of a representation of the Poincare group
\cite{Sav01a}. This Poincare group acts as
\begin{equation}
\hat{\phi}_n(X) \mapsto \hat{\phi}_{\Lambda n} (\Lambda X)
\end{equation}
and generates changes in the space-time foliation. Heisenberg
picture fields can be defined using the time-averaged Hamiltonian
which is also foliation dependent. These Heisenberg picture
fields are of the form $\hat{\phi}_n(X,s)$, and there is a second
representation of the Poincare group in which the boosts act in
the `internal' time direction s, and leave the foliation fixed.

\smallskip

In a previous paper \cite{Nol01a} we considered the extension of
the classical analogue of the above theory to the case of
electrodynamics. It was argued that in order to preserve the two
representations of the Poincare group, the history fields should
have \emph{five} components as opposed to the usual four. The
extra component is associated with the internal time direction.
It is important to note that the theory is not covariant under
the action of the $SO(3,2)$ isometry group of the space-time
manifold $\M \times \Real$, but only under the internal and
external $SO(3,1)$ subgroups. It was also shown in \cite{Nol01a}
how the extra component leads to an extra constraint, and thus to
an enlarged gauge group.

The theory in \cite{Nol01a} is concerned with history
configuration fields $A_M^n(X)$, and their canonical momenta
$E^M_n(X)$. The index $M$ runs from $0$ to $4$, $X$ is a
four-vector and $n$ is a future-pointing time-like unit vector
which labels a particular foliation of space-time. These fields
can be considered as vectors tangent to the space of embeddings
of $\M$ into $\N \simeq \M \times \Real$. The history fields
satisfy the Poisson algebra
\begin{eqnarray}
\label{covalg}
\{ A^n_M(X) , A^n_N(Y) \} &=& 0 \\
\{ E_n^M(X) , E_n^N(Y) \} &=& 0 \\
\{ A^n_M(X) , E_n^N(Y) \} &=& \delta^N_M \delta^{(4)}(X-Y).
\end{eqnarray}
The four-vector $n$ can be embedded in $\N$, resulting in a
five-vector $\tilde{n}^M$ given in coordinates by $(n,0)$. We
also have the five-vector $\tilde{e}^M$ given by $(0,1)$, and we
use these vectors to decompose the fields into their temporal
components
\begin{equation}
A^n_t(X) := \tilde{n}^M A^n_M(X) \,\,\, , \,\,\,  A^n_s(X) :=
\tilde{e}^M A^n_M(X)
\end{equation}
and similarly for the momentum field. The $n$-spatial projection
tensor is defined as
\begin{equation}
{}^nP^M_N = \delta^M_N - \tilde{e}^M \tilde{e}_N - \tilde{n}^M
\tilde{n}_N,
\end{equation}
and can be used to decompose fields on the five-dimensional
space-time into their `$n$-spatial' components
\begin{equation}
{}^n A_M(X) := {}^nP^N_M \, A_N^n(X).
\end{equation}
We define `$n$-spatial tensors' in a similar way, \eg
\begin{equation}
\label{nspatialtensor} {}^nF_{MN}(X) := {}^nP^R_M \, {}^nP^S_N
F_{RS}^n(X).
\end{equation}
The first-class constraints can now be written as
\begin{eqnarray}
\label{constraint1} E_n^s (X) &\approx& 0 \\
\label{constraint2} E_n^t (X) &\approx& 0 \\
\label{constraint3} {}^nE^M_\parallel(X) &\approx& 0,
\end{eqnarray}
where the longitudinal component of the electric field is defined
by
\begin{equation}
{}^nE_\parallel(X) := {}^n \p_M \, {}^nE^M (X).
\end{equation}

\subsection{Reduced State Space}

We now augment the constraints with the histories radiation gauge
conditions:
\begin{eqnarray}
\label{radgauge1} A^n_s (X) &=& 0 \\
\label{radgauge2} A^n_t (X) &=& 0 \\
\label{radgauge3} {}^nA_\parallel(X) &=& 0
\end{eqnarray}
where
\begin{equation}
{}^nA_\parallel(X) := \frac{{}^n\p^M \, {}^nA_M (X)}{\triangle_n}
\end{equation}
and $\triangle_n = {}^n \p_M \, {}^n \p^M$. The six equations
(\ref{constraint1}) - (\ref{constraint3}) and (\ref{radgauge1}) -
(\ref{radgauge3}) form a second class set of constraints and we
can follow the usual procedure to find the Dirac brackets of the
reduced history space $\C_n$. In terms of the transverse fields
\begin{eqnarray}
{}^nA^\perp_M(X) &:=& {}^nA_M(X) - {}^n\p_M {}^nA^\parallel(X) \\
{}^nE_\perp^M(X) &:=& {}^nE^M(X) - {}^n\p^M
\frac{{}^nE_\parallel(X)}{\triangle_n},
\end{eqnarray}
they turn out to be
\begin{eqnarray}
\{ {}^nA^\perp_M(X) , {}^nA^\perp_N(Y) \}_D &=& 0 \\
\{ {}^nE_\perp^M(X) , {}^nE_\perp^N(Y) \}_D &=& 0 \\
\{ {}^nA^\perp_M(X) , {}^nE_\perp^N(Y) \}_D &=& ({}^n P_M^N -
\triangle_n^{-1} \, {}^n\p_M \, {}^n\p^N)\delta^{(4)}(X-Y).
\end{eqnarray}
The right hand side of this algebra has become explicitly
foliation dependent with the non-covariant gauge choice. In the
radiation gauge the time-averaged Hamiltonian is
\begin{equation}
H^0_n = \int d^4X (\frac{1}{2} \, {}^n E^\perp_M \, {}^n
E_\perp^M + \frac{1}{4} \, {}^n F_{MN} \, {}^n F^{MN}),
\end{equation}
where $F^n_{MN} = 2 \p_{[M} A^n_{N]}$ and ${}^nF_{MN}$ is the
corresponding $n$-spatial tensor (\emph{cf}. equation
(\ref{nspatialtensor})).

\subsection{Quantisation}

We wish to find an irreducible representation of the commutator
algebra
\begin{eqnarray}
\label{redalg1}
\left[{}^n\hat{A}^\perp_M(X) , {}^n\hat{A}^\perp_N(Y)\right] &=& 0 \\
\label{redalg2}
\left[{}^n\hat{E}_\perp^M(X) , {}^n\hat{E}_\perp^N(Y)\right] &=& 0 \\
\label{redalg3} \left[{}^n\hat{A}^\perp_M(X) ,
{}^n\hat{E}_\perp^N(Y)\right] &=& i\hbar \, ({}^n P_M^N -
\triangle_n^{-1} \, {}^n\p_M \, {}^n\p^N)\delta^{(4)}(X-Y)
\end{eqnarray}
on a Hilbert space such that the radiation gauge quantum
Hamiltonian is represented by a \emph{self-adjoint} operator. The
self-adjointness condition is required to select one of the
infinitely many unitarily inequivalent representations of the
infinite dimensional algebra (\ref{redalg1}) - (\ref{redalg3}).
Such a representation exists on the bosonic Fock space $\E^{red}
:= \F_B[L^2(\Real^4)] \otimes \F_B[L^2(\Real^4)]$. This space is
associated with annihilation and creation operators which obey
the following algebra
\begin{eqnarray}
\left[ \hat{a}_a(X) , \hat{a}_b(X') \right] &=& 0 \\
\left[ \hat{a}_a(X) , \hat{a}^\dagger_b(X') \right] &=& \hbar
\delta_{ab} \delta^{(4)}(X - X'),
\end{eqnarray}
for $a = 1,2$. Using the Fourier transformed operators:
\begin{eqnarray}
\hat{a}^\dagger_a(K) &=& \frac{1}{(2\pi)^2} \int d^4X
\hat{a}^\dagger_a(X)
e^{iK \cdot X} \\
\hat{a}_a(K) &=& \frac{1}{(2\pi)^2} \int d^4X \hat{a}_a(X) e^{-iK
\cdot X},
\end{eqnarray}
we define field operators satisfying the algebra (\ref{redalg1}) -
(\ref{redalg3}) in the following way
\begin{eqnarray}
{}^n\hat{A}^\perp_M(X) = \frac{1}{(2\pi)^2} \sum_{a = 1}^2 \,
\int \frac{d^4K}{\sqrt{2\omega_n(K)}} \, {}^n\epsilon^a_M(K) [
\hat{a}_a(K)
e^{-iK \cdot X} + \hat{a}^\dagger_a(K) e^{iK \cdot X} ]  \\
{}^n\hat{E}_\perp^M(X) = \frac{1}{i(2\pi)^2} \sum_{a = 1}^2 \,
\int d^4K \sqrt{\frac{\omega_n(K)}{2}} \, {}^n\epsilon_a^M(K)
[\hat{a}_a(K) e^{-iK \cdot X} - \hat{a}^\dagger_a(K) e^{iK \cdot
X} ]
\end{eqnarray}
In the above expressions, $K$ is a four-vector representing the
four-momentum of a photon and $\omega_n(K)$ is the modulus of the
$n$-spatial four vector $(\delta^\mu_\nu - n^\mu n_\nu)K^\nu$
with respect to the Minkowski metric. For each $K$, we define the
five-vector $\tilde{K}$ to be the embedding of $K$ into the
five-dimensional space-time \N. In coordinates $(X,s)$ the vector
$\tilde{K}$ can be written as $(K,0)$. The five-vectors
${}^n\epsilon_a^M(K)$ are a pair of mutually orthogonal,
$n$-spatial unit vectors which are in addition orthogonal to the
vector ${}^n\tilde{K}^M = {}^nP^M_N \, \tilde{K}^N$. These
vectors satisfy the following completeness relation:
\begin{equation}
\sum_{a = 1}^2 {}^n\epsilon^a_M(K) \, {}^n\epsilon_a^N(K) = {}^n
P_M^N - \frac{{}^n\tilde{K}_M {}^n\tilde{K}^N}{\omega_n(K)^2}.
\end{equation}
This property ensures that the algebra defined by equations
(\ref{redalg1}) - (\ref{redalg3}) is satisfied. Using the fact
that the polarisation vectors are orthonormal
\begin{equation}
{}^n\epsilon_a^M(K)  \, {}^n\epsilon^b_M(K) = \delta_a^b,
\end{equation}
the normal-ordered time-averaged Hamiltonian can be written
\begin{equation}
\hat{H}^0_n = \sum_{a = 1}^2 \, \int d^4K \omega_n(K)
\hat{a}_a^\dagger(K) \hat{a}_a(K).
\end{equation}
This Hamiltonian generates translations in internal time, and it
is easy to see that
\begin{equation}
e^{-is\hat{H}^0_n/\hbar} \hat{a}_a(K) e^{is\hat{H}^0_n/\hbar} =
e^{is\omega_n(K)} \hat{a}_a(K).
\end{equation}
Following the argument in \cite{IshLin98}, these transformations
are unitarily implementable and we conclude that $\hat{H}^R_n$
exists as a self-adjoint operator in this representation.
Therefore, for each $n$, there exists a unitarily inequivalent
representation of the radiation gauge history algebra on the Fock
space $\E^{red}$.

\subsection{The Decoherence Functional}

In the case when history propositions are realised as the lattice
of projection operators on a Hilbert space $\V$, every decoherence
functional $d$ can be written in the form \cite{IshLin94}
\begin{equation}
d(\alpha,\beta) = Tr_{\V \otimes \V} (\alpha \otimes \beta \,
\Theta_d),
\end{equation}
where $\Theta_d$ is an operator on $\V \otimes \V$. In fact
$\Theta_d$ must satisfy certain conditions for $d$ to be a
decoherence functional \cite{IshLin94}. In the case of the scalar
field, the operator $\Theta$ is dependent on the foliation, and
can be written \cite{Sav01a},
\begin{equation}
\Theta_n = \langle 0 | \rho_{-\infty} | 0 \rangle
(\cS\U)_n^\dagger \otimes (\cS\U)_n.
\end{equation}
The quantum history space for the scalar field is the bosonic
Fock space $\F_B[L^2(\Real^4)]$. To each operator $O$, on the
base space $L^2(\Real^4)$, there is an associated operator on
$\F_B[L^2(\Real^4)]$ defined by
\begin{equation}
\Gamma(O) = O \oplus (O \otimes O) \oplus \cdots .
\end{equation}
Using this construction, the operator $(\cS\U)_n$ can be written
\begin{equation}
(\cS\U)_n = \Gamma(1+i\sigma_n),
\end{equation}
where $\sigma_n = n^\mu \p_\mu + (-\triangle_n+m^2)^\frac{1}{2}$
is an operator on the base Hilbert space $L^2(\Real^4)$. The
operator $\sigma_n$ is related to the canonical history action
$S_n$ in the following simple way \cite{Sav01a}
\begin{equation} \label{actiondec}
e^{isS_n} = \Gamma(e^{is\sigma_n}).
\end{equation}
Equation (\ref{actiondec}) can be used to \emph{define} the
decoherence functional corresponding to a given action operator
$S$ on a Fock space $\F = \F_B[\h]$. Firstly $S$ defines $\sigma$,
the `generator' of the decoherence functional, which is the
operator on $\h$ given by
\begin{equation}
e^{isS} = \Gamma(e^{is\sigma}).
\end{equation}
Now the decoherence functional is defined by the operator on $\F
\otimes \F$ given by
\begin{equation}
\Theta_S = \langle 0 | \rho_{-\infty} | 0 \rangle
\Gamma(1+i\sigma)^\dagger \otimes \Gamma(1+i\sigma).
\end{equation}

The reduced canonical history action of electrodynamics in the
radiation gauge is
\begin{equation}
S^{red}_n = \int d^4X \, ({}^n\hat{E}^M_\perp \, \p_n^t \,
{}^n\hat{A}_M^\perp - \hat{H}^0_n),
\end{equation}
where $\p_n^t := \tilde{n}^M \p_M$.The corresponding generator of
the decoherence functional is the operator $\sigma^{red}_n$,
defined on vectors of the form $f \otimes g$ in the base Hilbert
space $\h^{red} = L^2(\Real^4) \otimes L^2(\Real^4)$ by
\begin{equation}
\sigma^{red}_n(f \otimes g) = (n^\mu \p_\mu -
\triangle_n^\frac{1}{2})f \otimes (n^\mu \p_\mu -
\triangle_n^\frac{1}{2})g,
\end{equation}
and extended to an operator on $\h^{red}$ by linearity. The
associated decoherence functional is denoted $d_n^{red}$. The pair
$(P(\E^{red}),d_n^{red})$ is the reduced history theory of pure
quantum electrodynamics with respect to the foliation $n$.

\section{BRST Cohomology}

As shown in the previous section, electrodynamics can be quantised
starting from the reduced state space. This is because the reduced
state space has a simple structure, in particular it is a linear
space. This is not the case for other constrained field theories
such as Yang-Mills theory and gravity. For such theories another,
more general, approach is needed. The BRST formalism
\cite{HennTeit} is a powerful approach to the quantisation of
constrained systems, and can be formulated using rigorous
operator methods. Motivated by these considerations we develop
the BRST approach to the quantum theory of histories
electrodynamics. In this section we follow closely the notation
of \cite{HennTeit}.

\subsection{Classical BRST Cohomology}

The central idea in the BRST formalism is to extend the state
space by including fermionic `ghost' fields. The extended state
space maintains manifest covariance and locality, unlike the
reduced state space approach. The BRST charge is constructed from
the ghost fields and the constraints and, in the classical case,
is a functional on the extended state space. The BRST charge
generates \emph{nilpotent} canonical transformations on the
extended state space. The physical degrees of freedom are
identified with the corresponding set of cohomology classes. The
idea is that the ghost fields cancel out the gauge fields in the
cohomology.

We begin by briefly recalling the BRST approach to the standard
classical theory of electrodynamics as given in chapter 19 of
reference \cite{HennTeit}. We have fields $E^\mu(\underline{x})$
and $A_\mu(\underline{x})$ satisfying the algebra
\begin{equation}
\{ A_\mu(\underline{x}) , E^\nu(\underline{x}') \} =
\delta_\mu^\nu\delta^{(3)}(\underline{x} - \underline{x}'),
\end{equation}
and a pair of constraints $E^0(\underline{x}) \approx 0$ and
$\p_iE^i(\underline{x}) \approx 0$. Corresponding to the first
constraint we add a ghost pair $\eta(\underline{x}) ,
\cP(\underline{x})$ with
\begin{equation}
\{ \cP(\underline{x}) , \eta(\underline{x}') \} = -
\delta^{(3)}(\underline{x} - \underline{x}'),
\end{equation}
where the bracket is \emph{symmetric} representing the fact that
the ghost fields are fermionic. The Lagrange multiplier field
$A_0(\underline{x})$ and its conjugate momentum are associated
with an antighost field $\bar{C}(\underline{x})$ and conjugate
momentum $\rho(\underline{x})$ satisfying
\begin{equation}
\{ \rho(\underline{x}) , \bar{C}(\underline{x}') \} = -
\delta^{(3)}(\underline{x} - \underline{x}').
\end{equation}
The BRST charge is
\begin{equation} \label{BRSTOmega}
\Omega = \int d^3\underline{x} \, [-i\rho \, E^0 + \eta \,
\p_iE^i],
\end{equation}
and $\Omega$ generates \emph{nilpotent} canonical transformations
which are explicitly given in \cite{HennTeit}, and we denote by
$\tau$, that is $\tau(F) := \{\Omega,F\}$.  A functional F is said
to be BRST-closed if and only if
\begin{equation}
\tau(F) = 0,
\end{equation}
and a functional $G$ is said to be BRST-exact if and only if
\begin{equation}
G = \tau(G'),
\end{equation}
for some functional $G'$. By the nilpotency of $\tau$ closed
functionals are exact, but the converse is not necessarily true,
and the set of functionals which are closed but not exact is
isomorphic to the set of functionals on the reduced state space.
In addition \cite{HennTeit}, there is a natural Poisson algebra
defined on the set of cohomology classes which is a Poisson
subalgebra of the extended Poisson algebra, and is isomorphic to
the Poisson algebra of the reduced classical history space.

This analysis is easy to extend to the classical history theory of
electrodynamics. Corresponding to the constraint
${}^nE_\parallel(X) \approx 0$ we introduce a pair of fermionic,
scalar ghost history fields $\eta^1_n(X)$ and $\cP^1_n(X)$ which
satisfy the algebra
\begin{equation}
\{ \cP^1_n(X) , \eta^1_n(X') \} = -\delta^{(4)}(X - X').
\end{equation}
In addition we have two Lagrange multipliers in the history
theory so we add two antighost fields $C^n_a(X)$ $(a \in
\{1,2\})$, along with their conjugate momenta $\rho_n^a(X)$. These
fields satisfy
\begin{equation}
\{ \rho_n^a(X) , C^n_b(X') \} = -\delta^a_b \delta^{(4)}(X - X').
\end{equation}
In this way the ring of functions on the extended classical
history space is given the structure of a graded Lie algebra. The
history BRST charge $\Omega'_n$ is defined as
\begin{equation}
\Omega'_n = \int d^4X [ -i \rho^1_n E^t_n -i \rho_n^2 E_n^s +
\eta^1_n \, {}^nE_\parallel ],
\end{equation}
and generates canonical transformations denoted by $\tau^n$. The
transformations are
\begin{eqnarray}
\tau^n ({}^nA^\parallel) = \eta_n^1, \,\,\,\,\,\, \tau^n
(\eta_n^1) &=& 0, \,\,\,\,\,\, \tau^n (\cP_n^1) =
{}^nE^\parallel, \,\,\,\,\,\, \tau^n ({}^nE^\parallel) = 0, \,\,\,\,\,\, \\
\tau^n (A^n_s) = -i\rho_n^1, \,\,\,\,\,\,  \tau^n (A^n_t) &=&
-i\rho_n^2, \,\,\,\,\,\, \tau^n (\rho_n^1) = 0, \,\,\,\,\,\,
\tau^n (\rho_n^2) = 0, \,\,\,\,\,\, \\
\tau^n (C_n^1) = iE_n^s, \,\,\,\,\,\, \tau^n (C_n^2) &=& iE_n^t,
\,\,\,\,\,\, \tau^n (E_n^s) = 0 \,\,\,\,\,\,
\tau^n (E_n^t) = 0 \,\,\,\,\,\,  \\
\tau^n ({}^nA^\perp) &=& 0, \,\,\,\,\,\, \tau^n ({}^nE_\perp) = 0.
\end{eqnarray}
From these transformations it is clear that $\tau^n$ is
nilpotent, and thus defines a cohomology on the space of
functionals on the BRST-extended classical history space. The
classical history BRST cohomology, $H_{cl}^*(\Omega'_n)$, is
defined to be the space of equivalence classes of BRST-closed
functionals modulo BRST-exact ones. The only fields which are
closed but not exact are ${}^nA^\perp$ and ${}^nE_\perp$, and so
the cohomology classes are in bijective correspondence with
functionals of the transverse fields. Thus $H_{cl}^*(\Omega'_n)$
is isomorphic to the space of functionals on the reduced
classical history space.

\subsection{Operator Quantisation}

The BRST operator quantisation of standard electrodynamics
proceeds by expanding the quantum fields in terms of operators
which satisfy the algebra of creation and annihilation operators,
thus defining a representation of the field algebra on a Fock
space. Then, using the quantum BRST charge $\hat{\Omega}$, which
is the operator corresponding to the functional in equation
(\ref{BRSTOmega}), a cohomology can be defined on operators on the
quantum Hilbert space as follows. An operator $\hat{O}$ is
defined to be \emph{BRST-closed} if and only if
\begin{equation}
[\hat{\Omega},\hat{O}] = 0,
\end{equation}
and an operator $\hat{Q}$ is \emph{BRST-exact} if and only if
\begin{equation}
\hat{Q} = [\hat{\Omega},\hat{W}]
\end{equation}
for some operator $\hat{W}$. Because $\hat{\Omega}$ generates
nilpotent transformations and the commutator satisfies the graded
Jacobi identity, a BRST-exact operator is necessarily BRST-closed.
However, the converse is not true and two BRST-closed operators
$\hat{O}$ and $\hat{O}'$ are defined to be BRST-equivalent if
$\hat{O}' = \hat{O} + \hat{Q}$ for some BRST-exact operator
$\hat{Q}$. The \emph{operator cohomology} of $\hat{\Omega}$,
$H^*_{op}(\hat{\Omega})$, is defined to be the set of equivalence
classes of closed operators modulo this equivalence relation. The
expansion of the fields in terms of creation and annihilation
operators is given in \cite{HennTeit} and it follows that the
quantum BRST charge can be written in the form
\begin{equation} \label{quartetform}
\hat{\Omega} = \int d^3\underline{k} \,
[\hat{c}^\dagger(\underline{k})\hat{a}(\underline{k}) +
\hat{a}^\dagger(\underline{k})\hat{c}(\underline{k})].
\end{equation}
Now the non-physical field modes `cancel out' the ghost modes in
the cohomology or, more precisely, the operator cohomology is
isomorphic to the set of operators on the reduced quantum Hilbert
space. This is a consequence of a general result known as the
`quartet mechanism' \cite{HennTeit} which applies to any quantum
theory in which the BRST operator is the sum of terms of the form
(\ref{quartetform}).

\subsection{Quantum History Theory in Quartet Form}

For the Fock space quantisation of a BRST extended theory it is
necessary that the constraints come in pairs allowing the
definition of creation and annihilation operators as complex
linear combinations of pairs of fields. In the history theory
there are three constraints which cannot be grouped in pairs for
the Fock space quantisation. To proceed we include an extra
Lagrange multiplier along with the associated momenta and ghosts.
More precisely, we introduce a bosonic scalar field
$\hat{\lambda}_n(X)$, the Lagrange multiplier corresponding to
the constraint $\hat{E}_n^s(X) \approx 0$. Its canonical momentum
is denoted $\hat{B}_n(X)$ and is constrained to vanish
\begin{equation}
\hat{B}_n(X) \approx 0.
\end{equation}
The associated ghost pair is $(\hat{\eta}_n^2 , \hat{\cP}_n^2)$.
Now the fields do form `quartets', and the results of
\cite{HennTeit} can be applied. The detailed transformations of
the fields into the `quartet' form have been relegated to the
appendix. The important result is that the bosonic fields can be
defined in terms of six pairs of bosonic creation and
annihilation operators
\begin{eqnarray} \label{opquant1}
\left[ \hat{a}_a(K) , \hat{b}_b^\dagger(K') \right] &=&
-\hbar\delta_{ab}\delta^{(4)}(K-K') \\
\left[ \hat{b}_a(K) , \hat{a}^\dagger_b(K') \right] &=&
-\hbar\delta_{ab}\delta^{(4)}(K-K') \\
\label{opquant3} \left[ {}^\perp \hat{a}_a(K) , {}^\perp
\hat{a}^\dagger_b(K') \right] &=& \hbar\delta_{ab}
\delta^{(4)}(K-K'),
\end{eqnarray}
for $a , b \in \{1,2\}$. So we have a representation of the
bosonic part of the BRST-extended history algebra on the Fock
space $\E^B$ which is defined as the tensor product of six copies
of the bosonic Fock space $\F_B[L^2(\Real^4)]$. Similarly, the
fermionic fields can be expanded in terms of four pairs of
fermionic creation and annihilation operators, which satisfy the
anti-commutators
\begin{eqnarray} \label{opquantf1}
\left[ \hat{c}_a(K) , \hat{e}_b^\dagger(K') \right]
&=& -\hbar\delta_{ab}\delta^{(4)}(K-K') \\
\label{opquantf2} \left[ \hat{e}_a(K) ,\hat{c}_b^\dagger(K')
\right] &=& -\hbar \delta_{ab}\delta^{(4)}(K-K'),
\end{eqnarray}
where $a, b \in \{1,2\}$. In this way we have a representation of
the fermionic part of the BRST-extended history algebra on the
Fock space $\E^F$, which is defined as the tensor product of four
copies of the fermionic Fock space $\F_F[L^2(\Real^4)]$. The
whole algebra is represented on the space $\E = \E^B \otimes
\E^F$, which we call the BRST-extended quantum history space.

\subsection{Quantum Operator Cohomology}
The extended BRST charge is
\begin{equation}
 \label{Omegan}
\hat{\Omega}_n = \int d^4X \, [-i\hat{\rho}^1_n \hat{E}^s_n +
\hat{\eta}^1_n {}^n\hat{E}_\parallel -i\hat{\rho}^2_n \hat{E}^t_n
+ \hat{\eta}^2_n \hat{B}_n ],
\end{equation}
and in terms of oscillators it takes the $n$-independent form
\begin{equation}
\hat{\Omega} = \sum_{a=1}^2 \int d^4K [\hat{c}_a^\dagger(K)
\hat{a}_a(K) + \hat{a}_a^\dagger(K) \hat{c}_a(K)].
\end{equation}
The anti-Hermitian ghost number operator $\hat{\G}$ is defined in
terms of oscillators as
\begin{equation}
\hat{\G} = \sum_a \int d^4K(\hat{c}_a^\dagger(K) \hat{e}_a(K) -
\hat{e}_a(K)^\dagger \hat{c}_a(K)).
\end{equation}
Any operator $\hat{O}$ on \E \, can be decomposed in components of
definite ghost number
\begin{equation}
\hat{O} = \sum_g \hat{O}_g, \,\,\,\,\,\,
[\hat{\G},\hat{O}_g]=g\hat{O}_g \,\,\, , \,\,\, g \in \mathbb{Z}.
\end{equation}
It follows from these definitions that all bosonic fields are of
ghost number zero, the fields $\hat{\eta}^a$ and $\hat{\rho}^a$
are of ghost number $+1$, and $\hat{\cP}^a$ and $\hat{C}^a$ are
of ghost number $-1$. Vectors in the non-zero eigenspaces of
$\hat{\G}$ have an ill-defined scalar product, therefore a ghost
number zero condition is often imposed on the physical states.
However, in the Fock space quantisation the ghost number zero
condition is automatically satisfied in the cohomology classes.

 \smallskip

From equation (\ref{Omegan}) it is clear that for the case of
histories electrodynamics the operators $\hat{\eta}_n^a,
\hat{\rho}_n^a, {}^n\hat{E}_\parallel, \hat{E}_n^s, \hat{E}_n^t,
\hat{B}_n, {}^n\hat{A}^\perp, {}^n\hat{E}_\perp$ are closed.
Similarly, the operators $\hat{\eta}_n^a, \hat{\rho}_n^a,
{}^n\hat{E}_\parallel, \hat{E}_n^s, \hat{E}_n^t, \hat{B}_n$ are
exact; for example the smeared ghost field can be written
\begin{equation}
\hat{\eta}_n^1(f) =  [ \hat{\Omega} ,  \hat{A}_n^\parallel(f)].
\end{equation}
The transverse field operators are closed but not exact.

\subsection{Projection Operator Cohomology}

In a history theory it is projection operators that appear in the
decoherence functional, and operators which are not projectors
lack a direct physical interpretation. In the equivalence classes
of $H^*_{op}(\hat{\Omega})$, projection operators are identified
with operators which are not projection operators. This
identification is unnatural from the histories perspective so in
this subsection we define an equivalence relation directly on the
lattice of projectors. We then use this equivalence relation to
define $H^*(\hat{\Omega})$, the \emph{projection operator
cohomology} associated with $\hat{\Omega}$. Finally we show that
$H^*(\hat{\Omega})$ can be given the structure of a lattice, and
that this lattice is isomorphic to the lattice of projection
operators on the reduced quantum history space.
\begin{defn}
Given two closed projectors $\alpha$ and $\beta$, we say that
`$\alpha$ is an \emph{exact fine-graining} of $\beta$', written
$\alpha \preceq \beta$, if and only if $\beta = \alpha + \gamma$
for some projection operator $\gamma$ which is exact (\ie \,
$\gamma = [\hat{\Omega} , \hat{Q}]$ for some operator $\hat{Q}$)
and disjoint to $\alpha$.
\end{defn}
The relation $\preceq$ is a partial order. The \emph{primitive}
part of $\alpha$ is denoted $\alpha_0$ and is defined as the limit
of exact fine-grainings of $\alpha$. A unique $\alpha_0$ exists
for each closed $\alpha$ because $\preceq$ is a partial order. If
$\alpha$ is exact then $\alpha_0$ is the zero projector. If
$\alpha$ is closed but not exact then $\alpha_0$ projects onto the
spectrum of a closed but not exact field operator. For histories
electrodynamics we have seen that the closed but not exact field
operators are the transverse field operators, and so closed
primitive projectors are in bijective correspondence with elements
of $P(\E^{red})$.
\begin{defn}\label{projcoh}
Two BRST-closed projection operators $\alpha$ and $\beta$ are
said to be \emph{BRST-equivalent} if and only if $\, \alpha_0 =
\beta_0$.
\end{defn}
The projection operator cohomology $H^*(\hat{\Omega})$ is defined
as the space of closed projection operators modulo this
equivalence relation, and elements of $H^*(\hat{\Omega})$ are
identified with \emph{physical} propositions. Given a primitive
projection operator $\alpha_0$, the equivalence class containing
$\alpha_0$ is the collection of all exact coarse-grainings of
$\alpha_0$. The statement `$\alpha$ is an \emph{exact
fine-graining} of $\beta$' is equivalent to the statement `$\beta$
is an \emph{exact coarse-graining} of $\alpha$'. Let $[\alpha]$
denote the equivalence class containing the closed projector
$\alpha$. The map $\pi : [\alpha] \mapsto \alpha_0$ is
well-defined on $H^*(\hat{\Omega})$, and is in fact an isomorphism
between $H^*(\hat{\Omega})$ and $P(\E^{red})$; the inverse is
given by $\pi^{-1}: \alpha_0 \mapsto [\alpha_0]$. In order to give
$H^*(\hat{\Omega})$ the structure of a lattice, we need to examine
the geometry of the linear subspaces associated to closed and
exact projection operators.
\begin{defn}
A BRST-closed subspace $L \subset \E$ is a topologically closed
linear subspace of $\mbox{Ker}(\hat{\Omega})$.
\end{defn}
\begin{prop}
BRST-closed projection operators are in bijective correspondence
with BRST-closed subspaces of $\E$.
\end{prop}
\textbf{Proof}:\\
Each projection operator $\alpha$ is associated with a
topologically closed linear subspace $L_\alpha \subset \E$. If
$\alpha$ is a BRST-closed projection operator, \ie \,
$[\hat{\Omega},\alpha] = 0$, then by writing $\alpha$ in Dirac
notation as
\begin{equation}
\alpha = \sum_i | l_i \rangle \langle l_i |,
\end{equation}
where $| l_i \rangle$ is a basis of $L_\alpha$, it follows from
the independence of the basis vectors and the hermiticity of
$\hat{\Omega}$ that $\hat{\Omega} | l \rangle = 0$ for all $| l
\rangle \in L_\alpha$. Therefore $L_\alpha \subset
\mbox{Ker}(\hat{\Omega})$.

Conversely, each BRST-closed subspace is associated with a
BRST-closed projection operator because $\hat{\Omega}$ is
self-adjoint.
$\square$ \\
In the case of histories electrodynamics BRST-closed subspaces
are spanned by vectors created by the operators ${}^\perp
\hat{a}^\dagger_a , \hat{a}_a^\dagger, \hat{c}_a^\dagger$ acting
on the cyclic vacuum state.
\begin{defn}
A BRST-exact subspace $M \subset \E$ is a topologically closed
linear subspace of $\mbox{Im}(\hat{\Omega})$.
\end{defn}
\begin{prop}
BRST-exact projection operators are in bijective correspondence
with BRST-exact subspaces of $\E$.
\end{prop}
\textbf{Proof}:\\
A BRST-exact projection operator can be written in the form
$\gamma = [\hat{\Omega},\hat{Q}]$ for some operator $\hat{Q}$.
$\gamma$ is closed, and so is associated with a BRST-closed
subspace $M_\gamma \subset \E$. Now
\begin{equation}
\gamma | m \rangle = [\hat{\Omega} , \hat{Q}] | m \rangle =
\hat{\Omega}(\hat{Q}| m \rangle) \,\,\, \forall \,\, | m \rangle
\in M_\gamma
\end{equation}
because $\hat{\Omega} | m \rangle = 0$. However, $\gamma | m
\rangle = | m \rangle$ for any $| m \rangle \in M_\gamma$ so
\begin{equation}
| m \rangle = \hat{\Omega} (\hat{Q}| m \rangle) \,\,\, \forall
\,\, | m \rangle \in M_\gamma,
\end{equation}
and hence $M_\gamma \subset \mbox{Im}(\hat{\Omega})$.

Conversely, each BRST-exact subspace $M$ with basis $| m_i
\rangle$ is associated with an exact projection operator $\gamma$
\begin{equation}
\gamma = \sum_i |m_i \rangle \langle m_i| = [ \hat{\Omega} ,
\sum_i |u_{m_i} \rangle \langle m_i| ],
\end{equation}
where $| u_{m_i} \rangle$ is any vector such that $\hat{\Omega} |
u_{m_i} \rangle = | m_i \rangle$. $\square$ \\ In the case of
histories electrodynamics BRST-exact subspaces are spanned by
vectors created by the operators $\hat{a}_a^\dagger,
\hat{c}^\dagger_a$ on the vacuum state.
\begin{defn}
A primitive subspace $R \subset \E$ is a BRST-closed subspace
with no BRST-exact proper subspaces. The closure of the union of
all primitive subspaces is denoted $\E_0$.
\end{defn}
\begin{prop}
Primitive projection operators are in bijective correspondence
with primitive subspaces of $\E$.
\end{prop}
\textbf{Proof}:\\
Let $\alpha_0$ be a primitive projector. Then the only exact fine
graining of $\alpha_0$ is $\alpha_0$ itself. This implies that
the linear subspace associated with $\alpha_0$ has no BRST-exact
proper subspaces.

Conversely, because a primitive subspace has no BRST-exact proper
subspaces it follows that the corresponding projection operator
must be a limit of exact fine-grainings, and thus primitive.
 $\square$ \\
In the case of histories electrodynamics primitive subspaces are
spanned by vectors created by the action of the transverse
creation operators ${}^\perp \hat{a}^\dagger_a$ on the vacuum
state.

\medskip

From the above discussion it follows that $\E_0$ and
$\mbox{Im}(\hat{\Omega})$ are disjoint, and that the closure of
$\E_0 \cup \mbox{Im}(\hat{\Omega})$ is $\mbox{Ker}(\hat{\Omega})$.
Therefore every exact projector is disjoint to every primitive
projector, and $\mbox{id}_0 + \mbox{id}_{Im} = \mbox{id}_{Ker}$,
where $\mbox{id}_0, \mbox{id}_{Im}$ and $\mbox{id}_{Ker}$ are the
identity operators on $\E_0, \mbox{Im}(\hat{\Omega})$ and
$\mbox{Ker}(\hat{\Omega})$ respectively. These results can be
used to prove the following theorem.

\begin{thm}
(i) The lattice $P(\mbox{Ker}(\hat{\Omega}))$ induces a lattice
structure on $H^*(\hat{\Omega})$ by
\begin{eqnarray}
\left[ \alpha \right] \wedge \left[ \beta \right] &:=&
\left[ \alpha \wedge \beta \right] \\
\left[ \alpha \right] \vee \left[ \beta \right] &:=&
\left[ \alpha \vee \beta \right] \\
\neg \left[ \alpha \right] &:=& \left[ \neg\alpha \right].
\end{eqnarray}
\end{thm}
(ii) The map $\pi$ is a lattice isomorphism of
$H^*(\hat{\Omega})$ and
$P(\E^{red})$.\\
\textbf{Proof}:\\
(i) We have to show that the definitions give the same results
when evaluated on different members of an equivalence class. We
define the maximal exact part of $\alpha$ to be the unique exact
projector $\gamma_\alpha$ such that $\alpha = \alpha_0 +
\gamma_\alpha$. Every exact subspace is orthogonal to every
primitive subspace so $\alpha \wedge \beta = \alpha_0 \wedge
\beta_0 + \gamma_\alpha \wedge \gamma_\beta$. The intersection of
two exact subspaces is exact so $\gamma_\alpha \wedge
\gamma_\beta$ is exact, and $[\alpha \wedge \beta] = [\alpha_0
\wedge \beta_0]$. In a similar way we have $[\alpha \vee \beta] =
[\alpha_0 \vee
\beta_0]$.\\
Finally consider $\neg \alpha = \mbox{id}_{Ker} - (\alpha_0
+\gamma_\alpha)$ which can be written $\neg \alpha = (\mbox{id}_0
- \alpha_0) + (\mbox{id}_{Im} - \gamma_\alpha)$. Now
$\mbox{id}_{Im} - \gamma_\alpha$ is exact, so $[\neg \alpha] =
[\mbox{id}_0 - \alpha_0]$. Similarly we have $\neg \alpha_0 =
(\mbox{id}_0 - \alpha_0) + \mbox{id}_{Im}$ so $[\neg \alpha_0] =
[\mbox{id}_0 - \alpha_0]
= [\neg\alpha]$.\\
(ii) It is straightforward to check that\\
(a) $\pi([\alpha] \wedge [\beta]) = \pi[\alpha] \wedge \pi[\beta]$.\\
(b) $\pi([\alpha] \vee [\beta]) = \pi[\alpha] \vee \pi[\beta]$.\\
(c) $\pi(\neg [\alpha]) = \neg \pi[\alpha]$, \\
where the lattice operations on the right-hand-side of the above
equations are defined in $P(\E^{red})$. $\square$

Thus the projection operator cohomology is isomorphic to the
lattice of projection operators on the reduced history space. In
order to show that the corresponding history theories are `the
same' we first investigate the space of decoherence functionals
on the BRST-extended quantum history space.

\section{Gauge-Fixed Decoherence Functionals}

To each gauge-fixed action operator on $\E$, there corresponds a
gauge-fixed decoherence functional. The gauge-fixed decoherence
functionals assign non-trivial values to propositions regarding
the gauge and ghost fields. However, as we shall see, each
gauge-fixed decoherence functional induces a well-defined
functional on $H^*(\hat{\Omega})$ in such a way that the resulting
history theory is equivalent to the reduced quantum history
theory.

\subsection{Radiation and Feynman Gauges}

We begin by giving two explicit examples of gauge-fixed
decoherence functionals. The radiation gauge corresponds to the
following canonical action
\begin{equation} \label{radAct}
 \hat{S}^R_n = \hat{V}_n -  \hat{H}^0_n,
 \end{equation}
where the Louiville operator is the sum of three parts: \\
 1) A gauge-invariant part $\hat{V}_n^0$
 \begin{equation}
 \hat{V}_n^0 = \int d^4X \, {}^n\hat{E}^M_\perp \p_n^t \,
 {}^n\hat{A}_M^\perp,
 \end{equation}
 2) A ghost part
 \begin{equation}
 \hat{V}_n^{gh} = \int d^4X \sum_{a=1}^2 (\hat{\cP}_n^a \p_n^t \hat{\eta}_n^a +
 \hat{\rho}_n^a \p_n^t \hat{C}_n^a),
 \end{equation}
 3) A gauge part
 \begin{equation}
 \hat{V}_n^{ga} = \int d^4X ( \hat{E}_n^t \p_n^t \hat{A}^n_t + \hat{E}_n^s \p_n^t \hat{A}^n_s
 + {}^n\hat{E}^M_\parallel \p_n^t \,
 {}^n\hat{A}_M^\parallel + \hat{B}_n \p_n^t \hat{\lambda}_n).
 \end{equation}
The gauge-invariant Hamiltonian $\hat{H}_n^0$ is given by
\begin{equation}
\hat{H}_n^0 = \int d^4X (\frac{1}{2}{}^n\hat{E}^\perp_M \,
{}^n\hat{E}_\perp^M + \,
\frac{1}{4}{}^n\hat{F}_{MN}\,{}^n\hat{F}^{MN}).
\end{equation}
Using an argument similar to that in section 2, it follows that
the normal ordered Hamiltonian exists as a self-adjoint operator
on \E. The Louiville operator also exists, and therefore so does
the radiation gauge action $\hat{S}_n^R$. If the gauge and ghost
fields vanish initially, then they vanish identically on the
extrema of this action. However, note that the extrema of this
action satisfy the constraints if and only if the constraints are
satisfied initially.

\medskip

Let $\h_0 \simeq L^2(\Real^4) \otimes L^2(\Real^4)$ denote the
base Hilbert space of the primitive Fock space $\E_0$ \ie \, $\E_0
= \F_B[\h_0]$. Then the generator of the decoherence functional
associated with the radiation gauge action acts on vectors $f
\otimes g \in \h_0$ as
\begin{equation}
\sigma^R_n (f \otimes g) = (n^\mu \p_\mu -
\triangle_n^\frac{1}{2})f \otimes (n^\mu \p_\mu -
\triangle_n^\frac{1}{2})g,
\end{equation}
and is extended to an operator on $\h_0$ by linearity. The
radiation gauge Hamiltonian commutes with the gauge and ghost
fields. Therefore, as argued in \cite{Nol00}, projectors onto the
gauge and ghost fields form a canonical consistent set and the
probabilities assigned to these projectors is just the probability
in the initial state. This implies that if either $\epsilon$ or
$\kappa$ are projectors onto subspaces of the orthogonal
complement of $\E_0$, and the initial density matrix contains no
gauge or ghost modes, then
\begin{equation} \label{raddecfun}
d^R(\epsilon , \kappa) = 0.
\end{equation}
This completes the definition of $d^R$.

\medskip

In the `Feynman gauge' all fields satisfy the wave equation
internally. Let $\h$ denote the base Hilbert space of the
BRST-extended quantum history space $\E$. A vector in $\h$ is a
linear combination of homogeneous vectors of the form
\begin{equation}
v = \otimes_{i=1}^{10} v_i
\end{equation}
where $v_i \in L^2(\Real^4)$. The generator of the Feynman gauge
decoherence functional is defined on homogeneous vectors by
\begin{equation}
\sigma^F_n v = \otimes_{i=1}^{10} (n^\mu \p_\mu -
\triangle_n^\frac{1}{2}) v_i,
\end{equation}
and is extended to an operator on $\h$ by linearity.

\subsection{Gauge Transformations}

In this subsection we investigate gauge transformations. For
simplicity we consider the constraint $\hat{E}_n^t \approx 0$,
but analogous remarks apply to the other three constraints.

In the operator cohomology, the constraint $\hat{E}_n^t \approx
0$ is identified with operators of the form
\begin{equation}
\hat{D}_n^t = \hat{E}_n^t + [\hat{\Omega},\hat{Q}].
\end{equation}
The operator $\hat{D}_n^t$ generates gauge transformations in the
gauge field $\hat{A}^n_t$, and also in the ghost fields. We
require that the the ghost number zero eigenspace of $\hat{\G}$ is
mapped into itself under gauge transformations, which implies
that $[ \hat{\G} , \hat{D}_n^t ] = 0$. As $\hat{\Omega}$ is a
ghost number one operator, the operator $\hat{Q}$ must be of ghost
number $-1$. In addition we choose $\hat{Q}$ to be self-adjoint,
so that $\hat{D}_n^t$ generates unitary transformations.

$\hat{D}_n^t$ is exact, and can be written as
$[\hat{\Omega},\hat{G}_n]$ where $\hat{G}_n = \hat{C}_n^1 +
\hat{Q}$. It follows from the Jacobi identity that
\begin{equation}
[\hat{D}_n^t,\hat{O}] = [[\hat{\Omega} , \hat{G}_n],\hat{O}] =
[[\hat{G}_n,\hat{O}],\hat{\Omega}],
\end{equation}
if $\hat{O}$ is a closed operator. Thus closed operators are
mapped into exact operators by infinitesimal gauge
transformations. Under finite gauge transformations, closed
operators transform as
\begin{equation}
\hat{O} \mapsto \hat{U} \hat{O} \hat{U}^\dagger = \hat{O} +
[\hat{\Omega},\hat{W}].
\end{equation}
where $\hat{W}$ is a ghost number $-1$ operator. Therefore, gauge
transformations act trivially on the equivalence classes of the
operator cohomology.

As the gauge transformations are unitary, they map projection
operators onto projection operators and, because $\E_0$ and
$\mbox{Im}(\hat{\Omega})$ are disjoint, exact self-adjoint
operators commute with primitive projectors. Therefore the action
of a gauge transformation on a BRST-closed projector, $\alpha =
\alpha_0 + \gamma_\alpha$ is
\begin{equation}
\alpha \mapsto \hat{U} \alpha \hat{U}^\dagger = \alpha_0 + \hat{U}
\gamma_\alpha \hat{U}^\dagger
\end{equation}
This shows that the gauge group acts trivially on
$H^*(\hat{\Omega})$, and that primitive projectors are
\emph{gauge-invariant}.

\medskip

There is also a natural unitary action of the gauge
transformations on the space of decoherence functionals. The
decoherence functional $d$ is associated with an operator on $\E
\otimes \E$, denoted $\Theta_d$. Gauge transformations act on
$\Theta_d$ as $\Theta_d \mapsto \Theta_{d\,'}$ where
$\Theta_{d\,'} = U \otimes U \Theta_d U^\dagger \otimes
U^\dagger$. As shown in \cite{Sch96}, the operator $\Theta_{d\,'}$
is indeed associated with a \emph{bone fide} decoherence
functional $d'$. We say that the decoherence functionals $d$ and
$d'$ are \emph{related by the gauge transformation $U$}.
\begin{defn}
The collection of all decoherence functionals related to the
radiation gauge decoherence functional ,$d^R$, by a gauge
transformation is denoted $\D_{gf}$ and called the space of
\emph{gauge-fixed decoherence functionals}.
\end{defn}

\section{Gauge Invariance}

We now fix a particular foliation, drop the $n$-label, and use
coordinates adapted to the foliation. Let $\D$ denote the space of
decoherence functionals associated with the quantum history space
$\E$. A physical symmetry of a \emph{history} quantum theory
(PSHQT) realised on the Hilbert space $\E$ is defined in
\cite{Sch96} as an affine one-to-one map
\begin{eqnarray}
P(\E) \otimes P(\E) \times \D &\rightarrow&
P(\E) \otimes P(\E) \times \D \\
 \label{PSHQT}
([\alpha \otimes \beta] , \Theta) &\mapsto& ([\alpha \otimes
\beta]' , \Theta'),
\end{eqnarray}
that preserves the pairing between history propositions and
operators associated with decoherence functionals, \ie
\begin{equation}
 \label{pairing}
tr_{\E \otimes \E}([\alpha \otimes \beta] \Theta) = tr_{\E \otimes
\E}([\alpha \otimes \beta]' \Theta').
\end{equation}
Schreckenberg \cite{Sch96} proved the following histories analogue
of Wigners theorem:
\begin{thm}
Every PSHQT can be induced by a unitary or anti-unitary operator
$\hat{U}$ on \E in the sense that each PSHQT can be written as
\begin{eqnarray}
 \label{HS1}
[\alpha \otimes \beta] &\mapsto& \hat{U} \otimes \hat{U}
[\alpha \otimes \beta] \hat{U}^\dagger \otimes \hat{U}^\dagger \\
 \label{HS2}
\Theta &\mapsto& \hat{U} \otimes \hat{U} \Theta \hat{U}^\dagger
\otimes \hat{U}^\dagger,
\end{eqnarray}
for some unitary or anti-unitary operator $\hat{U}$. Conversely,
every transformation of the form (\ref{HS1}),(\ref{HS2}) for
unitary or anti-unitary $\hat{U}$ induces a PSHQT.
\end{thm}

We have seen that gauge transformations act unitarily on
projection operators and on the space of decoherence functionals.
Therefore gauge transformations induce a PSHQT and we can use the
histories analogue of Wigner's theorem. The following is an
immediate consequence:
\begin{prop} \label{prop}
For any gauge-fixed decoherence functional $d \in \D_{gf}$, and
any two primitive projectors $\alpha_0,\beta_0 \in P(\E_0)$,
\begin{equation}
d(\alpha_0,\beta_0) = d^{red}(\alpha_0,\beta_0).
\end{equation}
\end{prop}
\textbf{Proof}:\\
Since $d \in \D_{gf}$, it is related to $d^R$ by a gauge
transformation. We denote the unitary operator associated with
this gauge transformation by $\hat{U}_d$. The primitive projectors
$\alpha_0$ and $\beta_0$ satisfy $\hat{U}_d \alpha_0
\hat{U}_d^\dagger = \alpha_0$ and $\hat{U}_d \beta_0
\hat{U}_d^\dagger = \beta_0$. Now equation (\ref{pairing}) implies
$d(\alpha_0,\beta_0) = d^R(\alpha_0,\beta_0)$. Finally, from the
definition of $d^R$ it
follows that $d^R(\alpha_0,\beta_0) = d^{red}(\alpha_0,\beta_0)$. $\square$\\
This shows that gauge-fixed decoherence functionals ensure that:\\
(i) The \emph{probabilities} assigned to gauge-invariant
propositions are gauge-\\invariant.\\
(ii) The \emph{quantum interference} between gauge-invariant
propositions is gauge-invariant. \\
We have the following lemma:
\begin{lem}\label{lemma}
For any gauge-fixed decoherence functional \, $d \in \D_{gf}$, any
projector $\alpha \in P(\E)$, and any exact projector $\gamma$,
\begin{equation}
d(\alpha , \gamma) = 0.
\end{equation}
\end{lem}
\textbf{Proof}:\\
We act on $d(\alpha , \gamma)$ with $\hat{U}_d$, the gauge
transformation that maps $d$ into $d^R$. Under the action of
$\hat{U}_d$, $\alpha$ and $\gamma$ transform to $\alpha_d :=
\hat{U}_d \alpha \hat{U}^\dagger_d$ and $\gamma_d := \hat{U}_d
\gamma \hat{U}^\dagger_d$. Now using equation (\ref{pairing}) we
have
\begin{equation}
d(\alpha , \gamma) = d^R(\alpha_d , \gamma_d),
\end{equation}
which is equal to zero by equation (\ref{raddecfun}) because
$\gamma_d$ is exact.
 $\square$

\begin{cor}
Let $\gamma$ and $\delta$ be exact propositions. Then \,
$d(\gamma,\delta)=0$ for any gauge-fixed decoherence functional \,
$d \in \D_{gf}$.
\end{cor}
\textbf{Proof}: Immediate. $\square$\\
This implies that there is no interference between exact
projectors, and that exact propositions are assigned a probability
of zero by any gauge-fixed decoherence functional. Projectors onto
the spectrum of the constraint fields are exact, so a special case
of this result is that any gauge-fixed decoherence functional
assigns a probability of zero to any propositions that are not
compatible with the constraints.

\begin{thm} \label{mainthm}
Any gauge-fixed decoherence functional \, $d \in \D_{gf}$ reduces
to a well-defined functional $\tilde{d}: H^*(\hat{\Omega}) \times
H^*(\hat{\Omega}) \rightarrow \Complex$ defined by
\begin{equation}
\tilde{d}([\alpha] , [\beta]) := d(\alpha , \beta).
\end{equation}
In addition
\begin{equation} \label{historyiso}
\tilde{d}([\alpha] , [\beta]) = d^{red}(\alpha_0 , \beta_0),
\end{equation}
for all $[\alpha],[\beta] \in H^*(\hat{\Omega})$, where $\alpha_0$
and $\beta_0$ are the primitive parts of $\alpha$ and $\beta$
respectively.
\end{thm}
\textbf{Proof}: \\
We use the additivity axiom of the space of decoherence
functionals;
\begin{equation}
d(\alpha , \beta) = d(\alpha_0 + \gamma_\alpha , \beta) =
d(\alpha_0 , \beta) + d(\gamma_\alpha , \beta),
\end{equation}
because $\alpha_0$ and $\gamma_\alpha$ are disjoint. Now because
$\gamma_\alpha$ is exact, lemma (\ref{lemma}) along with the
hermiticity of $d$ implies that that $d(\gamma_\alpha , \beta) =
0$. This means that
\begin{equation}
d(\alpha , \beta) = d(\alpha_0 , \beta).
\end{equation}
Repeating this for the other argument of $d$, we have
\begin{equation}
d(\alpha,\beta) = d(\alpha_0 , \beta_0),
\end{equation}
for any gauge-fixed decoherence functional $d$. This shows that
every gauge-fixed decoherence functional reduces to a well-defined
functional $\tilde{d}: H^*(\hat{\Omega}) \times H^*(\hat{\Omega})
\rightarrow \Complex$ defined by
$\tilde{d}([\alpha],[\beta]) := d(\alpha,\beta)$. Now proposition (\ref{prop}) proves the theorem. $\square$\\
Theorem (\ref{mainthm}) shows that the cohomological history
theory $(H^*(\hat{\Omega}),\tilde{d})$ is `the same' as the
reduced history theory $(P(\E^{red}),d^{red})$. More precisely,
\begin{defn}
Two quantum history theories $(P_1 , \D_1)$ and $(P_2 , \D_2)$ are
defined to be \emph{isomorphic} if there exists (i) an isomorphism
of lattices $\lambda : P_1 \rightarrow P_2$, and (ii) a bijective
map $\vartheta : \D_1 \rightarrow \D_2$, such that
\begin{equation}
d(\alpha,\beta) = \vartheta(d)(\lambda(\alpha),\lambda(\beta)),
\end{equation}
for all $\alpha,\beta \in P_1$ and all $d \in \D_1$.
\end{defn}
\begin{prop}
$(H^*(\hat{\Omega}),\tilde{d})$ is isomorphic to
$(P(\E^{red}),d^{red})$.
\end{prop}
\textbf{Proof}\\
The map $\pi : [\alpha] \mapsto \alpha_0$ provides the required
lattice isomorphism between $H^*(\hat{\Omega})$ and
$P(\E^{red})$. Now define $\vartheta$ by $\vartheta(\tilde{d}) =
d^{red}$, and equation (\ref{historyiso}) states precisely that
the two history theories are isomorphic. $\square$

\section{Conclusion}

We have constructed two concrete models of history quantum
electrodynamics on Fock space. Firstly we quantised the classical
reduced history space by finding an inequivalent representation of
the reduced history algebra on $\E^{red}$ for each foliation. We
then defined the decoherence functional using the canonical
history action on the reduced history space. This results in the
reduced history theory $(P(\E^{red}) , d^{red})$.

Secondly we extended the history algebra by including ghost
fields, and found representations of the extended history algebra
on the extended quantum history space $\E$. Using the BRST charge
$\Omega$, we defined $H^*(\Omega)$, the projection operator
cohomology of $\Omega$, and showed that $H^*(\Omega)$ is
isomorphic (as a lattice) to $P(\E^{red})$. Finally we defined the
space of gauge-fixed decoherence functionals and showed that, for
each gauge-fixed decoherence functional $d$, $(H^*(\Omega) ,
\tilde{d})$ is isomorphic to $(P(\E^{red}) , d^{red})$.

\medskip

Although the construction of quantum history electrodynamics is
interesting in itself, it is hoped that the results obtained here
will be useful in a wider context. Given a general BRST-extended
quantum history space and a nilpotent BRST charge, section 3
provides a definition of the corresponding projection operator
cohomology, and shows that it is isomorphic to the lattice of
projection operators on the reduced history space. In addition,
the discussion of the space of gauge-fixed decoherence
functionals is relevant to any gauge theory. It would be
interesting to apply the histories BRST formalism developed here
to mini-superspace models, or to topological quantum field theory.
These examples are of particular importance in light of the
recent interest in diffeomorphism invariance in history theories
\cite{Sav01b,Kuch01}.

\section{Appendix}

Firstly we consider the bosonic part of the BRST-extended
commutator algebra:
\begin{eqnarray}
 \label{bosalg1}
\left[\hat{\lambda}_n (X) , \hat{B}_n(X')\right] &=& i\hbar\delta^{(4)}(X-X') \\
 \label{bosalg2}
\left[ \hat{A}^n_M(X) , \hat{E}_n^N(X') \right] &=&
i\hbar\delta^N_M \delta^{(4)}(X-X'),
\end{eqnarray}
where all unwritten commutators vanish. We expand the
configuration fields as
\begin{eqnarray}
\hat{\lambda}_n(X) &=& \frac{1}{(2\pi)^2} \int
\frac{d^4K}{\sqrt{2\omega_n(K)}} [\hat{d}(K)
e^{iK \cdot X} + \hat{d}^\dagger(K) e^{-iK \cdot X}] \\
\hat{A}^n_s(X) &=& \frac{1}{(2\pi)^2} \int
\frac{d^4K}{\sqrt{2\omega_n(K)}} [ \hat{d}_s(K)
e^{iK \cdot X} + \hat{d}^\dagger_s(K)  e^{-iK \cdot X} ] \\
\hat{A}^n_t(X) &=& \frac{1}{(2\pi)^2} \int
\frac{d^4K}{\sqrt{2\omega_n(K)}} [ \hat{d}_t(K)
e^{iK \cdot X} + \hat{d}^\dagger_t(K)  e^{-iK \cdot X} ] \\
{}^n\hat{A}_M(X) &=& \frac{1}{(2\pi)^2} \sum_{i=1}^3 \int
\frac{d^4K}{\sqrt{2\omega_n(K)}}  \\
&& [ \hat{d}_i(K) {}^n\epsilon_M^i(K) e^{iK \cdot X} +
\hat{d}^\dagger_i(K) {}^n\epsilon_M^i(K) e^{-iK \cdot X} ].
\end{eqnarray}
The five-vectors ${}^n\epsilon_M^a(K)$ for $a = 1,2$ are defined
as in section 2, and ${}^n\epsilon_M^3(K)$ is the unit vector
pointing in the direction of ${}^n\tilde{K}$. Now the momentum
fields are expanded as
\begin{eqnarray}
\hat{B}_n(X) &=& \frac{1}{i(2\pi)^2} \int d^4K \sqrt{\frac{\omega_n(K)}{2}} \\
&& [(\hat{d}(K) + \hat{d}_s(K)) e^{iK \cdot X} -
(\hat{d}^\dagger(K) +\hat{d}_s^\dagger(K)) e^{-iK \cdot X}] \\
\hat{E}^s_n(X) &=& \frac{1}{i(2\pi)^2} \int \, d^4K
\sqrt{\frac{\omega_n(K)}{2}} [ (\hat{d}_s(K)
+ \hat{d}(K)) e^{iK \cdot X} \\
&& - (\hat{d}^\dagger_s(K) + \hat{d}^\dagger(K))  e^{-iK \cdot X} ] \\
\hat{E}^t_n(X) &=& \frac{1}{i(2\pi)^2} \int \, d^4K
\sqrt{\frac{\omega_n(K)}{2}} [ (\hat{d}_t(K)
+ \hat{d}_3(K)) e^{iK \cdot X} \\
&& - (\hat{d}^\dagger_t(K) + \hat{d}^\dagger_3(K)) e^{-iK \cdot X} ] \\
{}^n\hat{E}^M(X) &=& \frac{1}{i(2\pi)^2} \sum_{i=1}^3 \int
\, d^4K \sqrt{\frac{\omega_n(K)}{2}} \\
&& [ (\hat{d}_i(K) {}^n\epsilon_M^i(K) +
\hat{d}_t(K) \, {}^n\epsilon_3^M(K)) e^{iK \cdot X} \\
&& - (\hat{d}^\dagger_i(K) \, {}^n\epsilon^M_i(K) +
\hat{d}^\dagger_t(K) \, {}^n\epsilon_3^M(K) )e^{-iK \cdot X} ].
\end{eqnarray}
Defining
\begin{eqnarray}
\hat{a}_1(K) = \frac{1}{\sqrt{2}} (\hat{d}_t(K) + \hat{d}_3(K)) \\
\hat{b}_1(K) = \frac{1}{\sqrt{2}} (\hat{d}_t(K) - \hat{d}_3(K)) \\
\hat{a}_2(K) = \frac{1}{\sqrt{2}} (\hat{d}_s(K) + \hat{d}(K)) \\
\hat{b}_2(K) = \frac{1}{\sqrt{2}} (\hat{d}_s(K) - \hat{d}(K)),
\end{eqnarray}
we obtain the commutators (\ref{opquant1})-(\ref{opquant3}) in
which $\hat{d}_a$ has been written ${}^\perp \hat{a}_a$ for $a
\in \{1,2\}$.

\medskip

The fermionic part of the BRST-extended quantum history algebra is
\begin{eqnarray}
 \label{feralg1}
\left[\hat{\eta}^a_n (X) , \hat{\cP}^b_n(X')\right] &=& \hbar
\delta^{ab}\delta^{(4)}(X-X') \\
 \label{feralg2}
\left[ \bar{C}^a_n(X) , \hat{\rho}^b_n(X') \right] &=& \hbar
\delta^{ab}\delta^{(4)}(X-X')
\end{eqnarray}
where on fermionic fields square brackets represent
anti-commutators. We write the ghost fields as
\begin{eqnarray}
\hat{\eta}^1_n(X) &=& -\frac{1}{(2\pi)^2} \int
\frac{d^4K}{2\omega_n(K)^{3/2}} [c_1(K) e^{i
K \cdot X} + c_1^\dagger(K) e^{-i K \cdot X} ] \\
\hat{\cP}^1_n(X) &=& \frac{i}{(2\pi)^2} \int d^4K \,
\omega_n(K)^{3/2}
[\bar{c}_1(K) e^{iK \cdot X} + \bar{c}_1^\dagger(K) e^{-i K \cdot X} ] \\
\hat{\rho}^1_n(X) &=& -\frac{1}{(2\pi)^2} \int
\frac{d^4K}{2\omega_n(K)^{1/2}} [c_1(K) e^{i
K \cdot X} - c_1^\dagger(K) e^{-i K \cdot X} ] \\
\bar{C}^1_n(X) &=& \frac{1}{i(2\pi)^2} \int d^4K \,
\omega_n(K)^{1/2} [\bar{c}_1(K)
e^{i K \cdot X} + \bar{c}_1^\dagger(K) e^{-i K \cdot X} ] \\
\hat{\eta}^2_n(X) &=& -\frac{1}{(2\pi)^2} \int
\frac{d^4K}{2\omega_n(K)^{1/2}} [c_2(K) e^{i
K \cdot X} + c_2^\dagger(K) e^{-i K \cdot X} ] \\
\hat{\cP}^2_n(X) &=& \frac{i}{(2\pi)^2} \int d^4K \,
\omega_n(K)^{1/2}
[\bar{c}_2(K) e^{iK \cdot X} + \bar{c}_2^\dagger(K) e^{-i K \cdot X} ] \\
\hat{\rho}^2_n(X) &=& -\frac{1}{(2\pi)^2} \int
\frac{d^4K}{2\omega_n(K)^{1/2}} [c_2(K) e^{i
K \cdot X} - c_2^\dagger(K) e^{-i K \cdot X} ] \\
\bar{C}^2_n(X) &=& \frac{1}{i(2\pi)^2} \int d^4K \,
\omega_n(K)^{1/2} [\bar{c}_2(K) e^{i K \cdot X} +
\bar{c}_2^\dagger(K) e^{-i K \cdot X} ]
\end{eqnarray}
and the algebra (\ref{feralg1}),(\ref{feralg2}) implies the
anti-commutators in equations (\ref{opquantf1}) ,
(\ref{opquantf2}).

\section{Acknowledgements}
I would like to thank Prof. Isham for suggesting this research and
for his comments on a draft of this paper. I am also grateful to
Dr N. Linden and Prof. B. Hiley for finding an error in a
previous version of this work. Financial support in the form of a
PPARC studentship is gratefully acknowledged.


\end{document}